\documentclass[12pt]{article}
\usepackage[latin1]{inputenc}
\usepackage{graphicx}
\newcommand{\mrm}{{\rm m}}

\newcommand{\text}{\rm}

\newcommand{\kr}{k_{\rho}}

\newcommand{\bb}{\begin{equation}}
\newcommand{\ee}{\end{equation}}
\newcommand{\bega}{\begin{eqnarray}}
\newcommand{\ega}{\end{eqnarray}}
\newcommand{\begae}{\begin{eqnarray*}}
\newcommand{\egae}{\end{eqnarray*}}

\newcommand{\h}{\hspace*{4ex}}

\newcommand{\om}{\omega}

\newcommand{\cent}{\centerline}
\newcommand{\vs}{\vspace*}

\begin{document}

\baselineskip 0.8cm

\

\begin{center}

{\large {\bf Proposte di Antenne generatrici di Fasci

Non-diffrattivi per Micro-onde

({\em Proposal of apertures generating Nondiffracting Beams

of Microwaves}) }}

\end{center}

\vs{3mm}

\cent{ Michel Zamboni Rached, }

\vs{0.2 cm}

\centerline{{\em DMO/FEEC, Universidade Estadual de Campinas,
SP, Brazil.}}

\vs{0.7 cm}

\cent{ Erasmo Recami }

\vs{0.2 cm}

\cent{{\em INFN---Sezione di Milano, Milan, Italy; \ {\rm and}}}
\cent{{\em Facolt\`a di Ingegneria, Universit\`a statale di Bergamo,
Bergamo, Italy.}}

\vs{0.2 cm}

\centerline{\rm and}

\vs{0.3 cm}

\cent{ Massimo Balma }

\vs{0.3 cm}

\cent{{\em Selexgalileo, Caselle (TO), Italy.}}

\vs{1. cm}

{\bf Abstract  \ --} \  We propose in detail Antennas for generating Nondiffracting Beams of Microwaves,
for instance with frequencies of the order of few GHz, obtaining fair results even when having recourse
to realistic apertures, with a quite reasonable diameter.  The present proposal refers to sets of
suitable annular slits. The possible applications are various, including remote sensing. The paper is
in Italian.  [Si propongono in dettaglio Antenne per la generazione
di fasci non-diffrattivi di microonde, per frequenze ad esempio
dell'ordine della decina di GHz, ottenendo discreti risultati pur ricorrendo ad antenne realistiche di
diametro ridotto.  La proposta \`e quella di usare un set
di opportuni Annular Slits. Le applicazioni possibili sono
varie, includendo il remore sensing.]\\

\vs{0.5 cm}

{\bf 1. -- Introduzione}\\

\h Per frequenze dell'ordine della decina di GHz si presenta il problema
di usare antenne di raggio non troppo elevato.

\h La frequenza ad esempio di 15 GHz corrisponde a una $\lambda$ di 2 cm, e quindi
anche un'antenna di diametro di 1.2 metri avrebbe un raggio  solo
30 volte superiore alla lunghezza d'onda. Ci\`o rende
abbastanza difficile l'ottenere fasci non-diffrattivi altamente
efficienti, vale a dire, con spots dell'ordine della lunghezza
d'onda $\lambda$, e con grande profondit\`a di campo. Onde
ottenere tali fasci, sarebbero necessarie aperture di difficile
realizzazione pratica, con raggio {\em molto}
maggiore di quello dello spot. Ci si propone di investiare la possibilit\'a,
ciononostante, di ottenere risultati buoni o comunque accettabili, anche
per il presente regime di frequenze, e limitandosi ad aperture di diametro
di circa 1 metro soltanto.

\h La nostra analisi si basa sull'approssimazione scalare. Un fascio di Bessel
(Bessel beam = Bb) con simmetria assiale pu\`o essere scritto:

\bb \psi(\rho,z,t) = J_0(\kr\rho)\exp{i(k_z z - \om t)}
\label{bb}\ee

\h In tale forma, il Bb \`e un fascio ``ideale", che si propaga
con una struttura trasversale di campo inalterata, e con uno spot dato da
$\Delta\rho = 2.4/\kr$ in qualsiasi sua posizione: diremo pertanto che
il fascio ideale possiede una profondit\`a di campo infinita.
Purtroppo il Bb ideale
abbisognerebbe di una apertura infinita, e quindi comporterebbe un
flusso di potenza infinito attraverso una superficie trasversale
quale la $z=0$.  E' necessario dunque  ``troncarlo".

\h Il detto Bb, quando viene troncato mediante una apertura finita di
raggio $R$ \ (tale che \ $R >> \Delta\rho$), \ passa a possedere una
profondit\`a di campo $Z$ finita, data da

\bb Z = R/\tan(\theta) \ee

dove $\theta$ \`e l'angolo di axicon del Bb, il quale angolo dipende
dai numeri d'onda  longitudinale e trasversale attraverso le relazioni
$k_z = \om/c\,\cos(\theta)$ \ e \ $\kr = \om/c\,\sin(\theta)$.

\h Nella regione $0<z<Z$ e $0<\rho<(Z-z)\tan(\theta)$, possiamo dire che
il Bb troncato  pu\`o essere ben approssimato dalla soluzione ideale (\ref{bb}).
Questo pu\`o essere dimostrato usando semplici argomenti di ottica geometrica. Per\`o,
quando l'apertura di troncamento possiede un raggio $R$ che non obbedisce alla
relazione $R>>\Delta\rho$, non si pu\`o pi\`u dire con certezza che il campo rimanga
non-diffrattivo nella detta regione, e ancor meno che in quella regione esso possa
essere approssimato dalla espressione del Bb ideale. In tali circostanze
diventa necessario ricorrere a simulazioni numeriche, basate sugli integrali di
diffrazione, per ottenere il campo emanato dall'apertura finita.

\h E questo \`e proprio il nostro caso.

\h Per accelerare il lavoro, useremo per\`o un nostro recente metodo teorico, in
corso di pubblicazione, capace di fornire espressioni {\em analitiche} anche per il
caso di un campo troncato. Tale metodo \`e abbastanza potente da averci permesso
di ottenere con pochi secondi quanto richiede interi giorni di calcolo con le usuali
simulazioni numeriche degli integrali di diffrazione. Esso \`e stato da noi
applicato per il caso, appunto, di antenne composte da aperture anulari.

\h Qui di seguito esponiamo in dettaglio la nostra proposta.

\

{\bf 2. -- Bessel beams con aperture finite}\\

\h Prima di suggerire il primo Prototipo, trattiamo delle
caratteristiche di un Bb troncato da una apertura
finita, ricorrendo al nostro nuovo metodo analitico.

\h Consideriamo un Bb con angolo di axicon $\theta = 0.062$ rad, frequenza 15 GHz
(e pertanto uno spot $\Delta\rho=12$ cm), troncato da una apertura circolare finita di
raggio $R=10$ m. \ Ci si aspetta in questo caso che il campo emanato sia approssimatamente
dato dalla (\ref{bb}) nella regione $0<z<Z$ e $0<\rho<(Z-z)\tan(\theta)$,
con $Z = 161.1$ m.

\h Si noti che per il Bb stiamo chiamando raggio dello spot la distanza, a
partire da $\rho=0$ (nella direzione trasversale), alla quale si trova il primo
zero della intensit\`a del campo. Si potrebbe adottare come raggio dello spot la distanza
dall'origine del punto ove la sua intensit\`a cade di un fattore $1/e$: in questo secondo caso
lo spot iniziale del Bb suddetto avrebbe un raggio $\Delta\rho(z=0)=7$ cm.

\h Nelle figure che seguono rappresentiamo: il campo all'apertura (Fig1a) con la
sua intensit\`a (Fig1b); nonch\'e l'intensit\`a 3D del campo
emanato (Fig1c) e la proiezione del suo andamento (Fig1d).

\begin{figure}[!h]
\begin{center}
 \scalebox{1}{\includegraphics{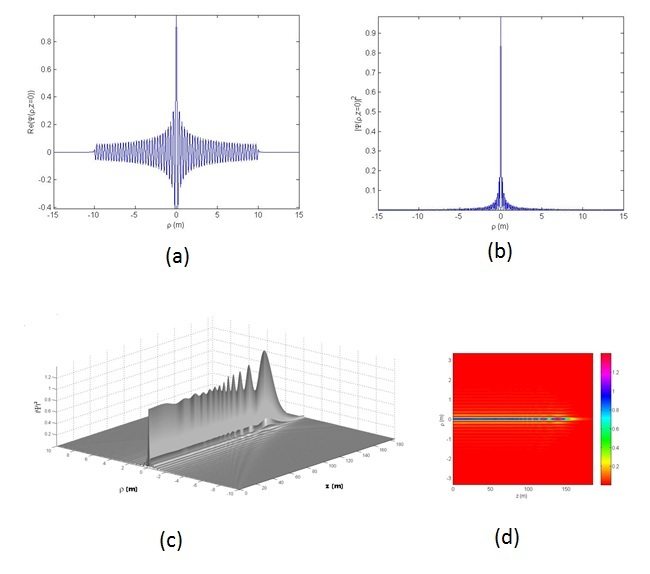}}
\end{center}
\caption{Vedere il testo.} \label{fig1}
\end{figure}

\newpage

\h Vediamo ora che succede quando il beam \`e troncato da una apertura molto pi\`u piccola, di
raggio ad esempio $R=61$ cm.  \ Se usassimo l'espressione $R/\tan(\theta)$
per la profondit\`a di campo, troveremmo il valore $Z=9.8$ m, \ ma dalle figure (Figs2a-2d)
si vede che il campo comincia a soffrire un intenso
decadimento (tipico dei fasci non-diffrattivi troncati) ad una distanza minore
(pi\`u o meno a $z=6$ m). E' possibile anche notare che gli anelli di intensit\`a laterale
(e ce ne sono solo tre in questo caso) cominciano a degradarsi ancor prima di questa
distanza: ci\`o succede proprio perch\'e i pochi anelli di intensit\`a non sono
capaci di ricostruire lo spot centrale alla (grande) distanza $Z$.

\begin{figure}[!h]
\begin{center}
 \scalebox{.9}{\includegraphics{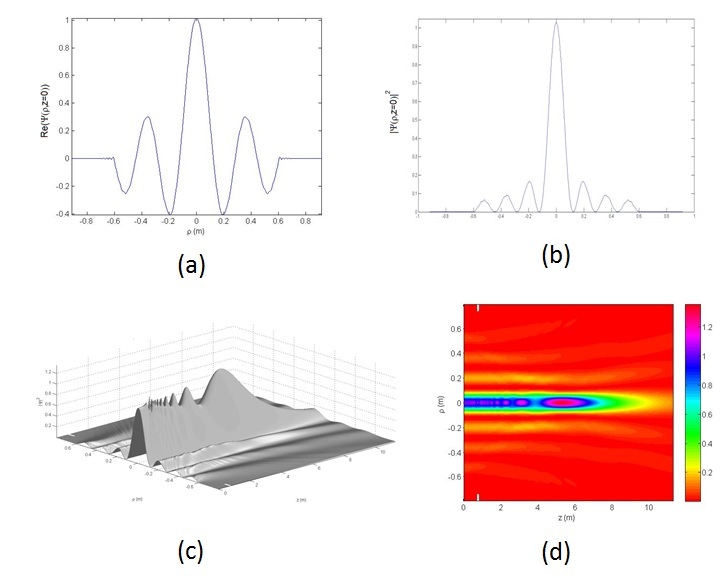}}
\end{center}
\caption{Figure mostranti il comportamento di un Bessel beam troncato da una apertura finita
piccola per le esigenze di un fascio non-diffrattivo. Con questa apertura, solo 3
anelli di intensit\`a sopravvivono al troncamento. } \label{fig2}
\end{figure}

\newpage

\h Ciononostante, notiamo che succede qualcosa di interessante. Bench\'e il fascio
inizi il suo decadimento prima di $Z=R/\tan(\theta)=9.8$ m, e pi\`u precisamente
a partire da $z=6$ m, la larghezza del suo spot {\em si mantiene} per distanze maggiori.

\h La figura che segue mostra l'andamento trasversale di intensit\`a in $z=0$ e dopo $10$ m
di propagazione, ossia in $z=10$ m. \ Vediamo che
l'intensit\`a dello spot decade di $1/4$ del suo valore iniziale, ma il suo raggio {\em si
altera poco}, da $\Delta\rho(z=0)=12$ cm  \ a \ $\Delta\rho(z=10 \; \mrm)=15$ cm  circa.

\begin{figure}[!h]
\begin{center}
 \scalebox{1.1}{\includegraphics{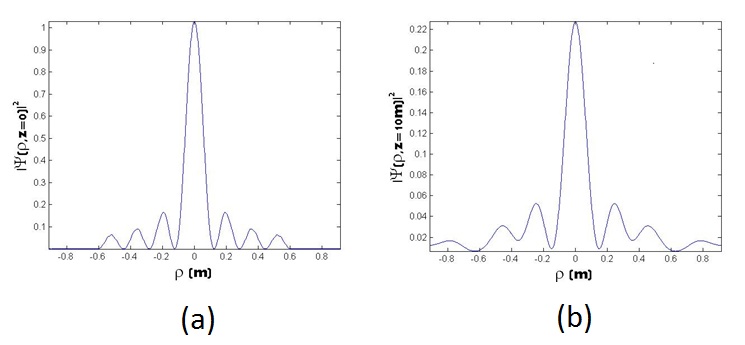}}
\end{center}
\caption{Confronto del comportamento trasversale di un Bb, troncato da una apertura di
raggio $R=61$ cm, rispettivamente: \ (a) nel piano dell'apertura, ovvero per
$z=0$;  e \ (b) dopo $10$ m di propagazione, ovvero in $z=10$ m. } \label{fig3}
\end{figure}

\h E' importante ricordare che un \textbf{fascio gaussiano} con uno spot iniziale di
raggio $\Delta\rho(z=0)=12$ cm raddoppierebbe la sua larghezza dopo
$3.9$ m, e in $z=10$ m il suo spot avrebbe intensit\`a centrale quasi sei volte minore
di quella iniziale,  e, soprattutto, avrebbe un raggio di $\Delta\rho(z=10 \; \mrm)=30$ cm,
quasi il triplo di quello iniziale.  Il set di figure che seguono mostra quanto si \`e
detto:

\

\begin{figure}[!h]
\begin{center}
 \scalebox{1.1}{\includegraphics{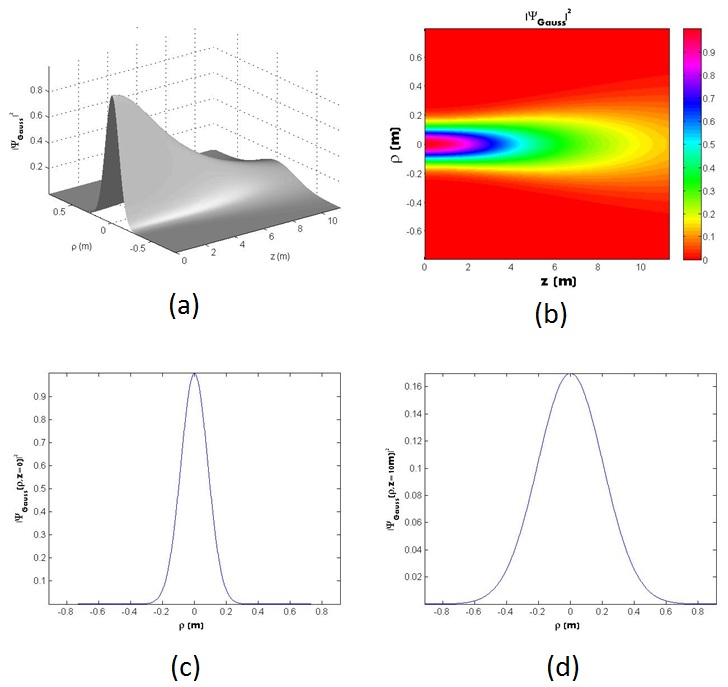}}
\end{center}
\caption{Figure mostranti l'evoluzione di un fascio {\bf gaussiano} con spot iniziale di
raggio $\Delta\rho=12$ cm} \label{fig4}
\end{figure}

\

\h {\em Possiamo pertanto concludere che}: bench\'e il {\bf Bessel} beam di cui sopra
sia fortemente troncato,
tanto da rimanere solo con tre dei suoi anelli di intensit\`a, esso \`e ancora capace
di mantenere la forma spaziale del suo spot (anche se non la sua intensit\`a) per
distanze {\em superiori} a quelle ottenute usando un fascio gaussiano.

\h Passiamo ora al nostro prototipo.

\

{\bf 3. -- Antenne composte da aperture anulari}\\

\h La nostra attuale proposta \`e semplice. Consideriamo una apertura circolare circondata da
un set di aperture anulari concentriche, e con tale array {\em tenteremo di riprodurre
in forma approssimata} un Bb con angolo di axicon $\theta = 0.062$ rad, frequenza $15$ GHz
(quindi con uno spot di $\Delta\rho=12$ cm), troncato da una apertura circolare finita di
raggio $R=61$ cm.

\h La nostra proposta \`e di modellare tale array di aperture anulari (pi\`u l'apertura
centrale), e l'eccitazione di ciascuna di esse, tenendo presente la forma stessa del
fascio di Bessel (troncato) desiderato. {\em Ossia, collochiamo gli slits tra gli zeri consecutivi
della funzione di Bessel in parola, ``illuminandoli" con campi uniformi che variano
di ampiezza (da uno slit all'altro) d'accordo con l'ampiezza del massimo modulo  della
funzione di Bessel nei rispettivi intervalli} (ci si riferisce, ripetiamolo, agli intervalli
tra gli zeri delle funzioni di Bessel, i quali individuano anche la posizione dei nostri
annular slits).

\h La figura che segue \`e auto-esplicativa:

\begin{figure}[!h]
\begin{center}
 \scalebox{1.1}{\includegraphics{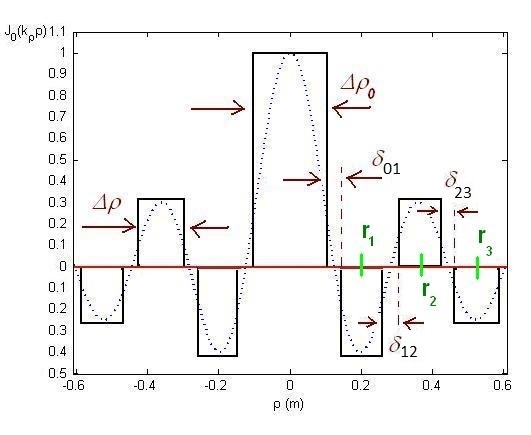}}
\end{center}
\caption{La figura mostra il modellamento spaziale dell'array di slits anulari, e le
sue rispettive eccitazioni, a partir dal Bb (troncato) desiderato, in $z=0$.
Le curve punteggiate indicano la funzione di Bessel scelta, e i gradini mostrano
posizioni e larghezze degli slits, cos\`{\i} come le ampiezze dei campi uniformi
da applicare a ciacuno di essi.} \label{fig5}
\end{figure}

\h Il nostro Prototipo preliminare \`e essenzialmente quello mostato in Figura $5$.
I valori dei parametri sono dati qui di seguito per tre casi:

\

\h \textbf{Prototipo numero 1}: $r_1 = \pi(1+1/4)/\kr = 0.20$ m, \ $r_2 = \pi(2+1/4)/\kr =
0.36$ m, \ $r_3 = \pi(3+1/4)/\kr = 0.52$ m, \ $\Delta\rho_0
= 0.23$ m, \ $\Delta\rho_1=\Delta\rho_2=\Delta\rho_2=\Delta\rho= 0.13$ m, \
$\delta_{01}=0.021$ m, \ $\delta_{12}=\delta_{23}=0.034$ m. \ \
I valori numerici delle ampiezze dei campi uniformi in ciascun slit sono dati
dai valori dei picchi della funzione di Bessel. Notare esplicitamente che le ampiezze
si alternano, passando da valori positivi a negativi, ad ogni cambiamento di slit:
ci\`o deve essere strettamente rispettato. Valori numerici dei campo negli
slits:\footnote{Qui $\Psi_n$ indica il valore numerico del campo nell'ennesimo slit, dove
$n=0$ si riferisce all'apertura circolare centrale (che possiamo chiamare slit numero zero).
Analogamente, $r_n$ ri riferisce al raggio dell'ennesimo slit,
mentre il raggio del circolo centrale \`e denominato $\Delta\rho_0$. Non si dimentichi
che $\kr = (\om/c) \sin (\theta) =  19.46 \mrm^{-1}$ } \ $\Psi_0 = 1$ a.u., \ $\Psi_1 = J_0(\kr r_1)
= -0.4026$ a.u., \ $\Psi_2 = J_0(\kr r_2) = 0.3001$ a.u., \ $\Psi_3 = J_0(\kr r_3) =
-0.2497$ a.u. \
%%, con $r_1 = \pi(1+1/4)/\kr$, $r_2 = \pi(2+1/4)/\kr $, $r_3 = \pi(3+1/4)/\kr$.
%% 0   -0.4026    0.3001   -0.2497

\h In Fig.6 mostriamo il campo emanato da una tale struttura (antenna).

\h Vedendo le Figs.6a e 6b, all'inizio si ha l'impressione che il fascio non assomigli a
quello di un Bessel troncato, ma questo \`e dovuto al fatto che in vicinanza dell'apertura
(antenna) il campo possiede alcuni picchi isolati di intensit\`a
(a causa dei tagli degli slits stessi) e questi picchi hanno l'effetto di rendere il campo
successivo visualmente poco nitido in Figura (principalmente nella Fig.6b, nella quale si
usano colori per indicare l'intensit\`a). Le Figs.6c e 6d mostrano il campo a partire da
$z=2.5$ m, e in esse si pu\`o avvertire la somiglianza con il campo di un Bessel troncato.
Nella Fig.6e il colore rosso mostra il campo all'apertura (antenna), ovvero in $z=0$,
mentre la linea punteggiata mostra la funzione di Bessel che viene ``discretizzata"
dai campi uniformi negli slits. {\em Nuovamente, \`e fondamentale comprendere che questa
Figura mostra la parte reale del fascio, coi suoi valori di ampiezza positivi e negativi.
Tali valori devono essere esattamente riprodotti nello schema per la generazione
del fascio.}  \ La Fig.6f mostra l'andamento dell'intensit\`a trasversale del fascio dopo
10 metri di propagazione, ossia in $z=10$ m.  Si vede che, nonostante la diminuizione
di intensit\`a (che cade di $1/3$ rispetto a quella che si ha in corrispondenza dell'antenna),
il valore del raggio dello spot muta molto poco.

\newpage
%%$\,$

\begin{figure}[!h]
\begin{center}
 \scalebox{1.05}{\includegraphics{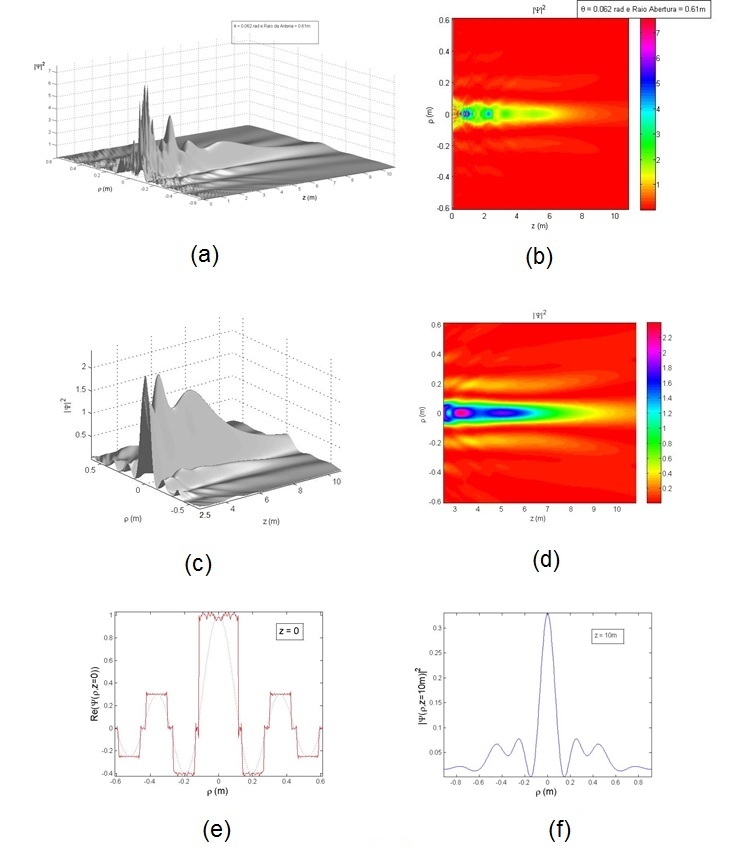}}
\end{center}
\caption{Campo emanato dal Prototipo 1. \ Spiegazioni nel testo.} \label{fig6}
\end{figure}

\newpage

\h \textbf{Prototipo numero 2}: Qui manteniamo inalterata tutta la struttura spaziale dell'antenna,
cambiando solo i valori dei campi uniformi che ``illuminano" ciascun slit.
Questo cambiamento \`e molto semplice: Non si muta il campo uniforme dell'apertura
circolare centrale, mentre negli slits ($n>1$) si moltiplica il
valore di ciascuna $\Psi_n$ per $\sqrt{n+1}$. \ Ovvero:

$\Psi_0 = 1$, \ $\Psi_1 = \sqrt{2}\;J_0(\kr r_1)=-0.57$ a.u., \
$\Psi_2 = \sqrt{3}\;J_0(\kr r_2)=0.52$ a.u., \ $\Psi_3 =
\sqrt{4}\;J_0(\kr r_3)=-0.5$ a.u..
%%0   -0.5694    0.5198   -0.4994

\h Le Fgure 7 mostrano il campo emanato dal Prototipo 2.

\begin{figure}[!h]
\begin{center}
 \scalebox{1.1}{\includegraphics{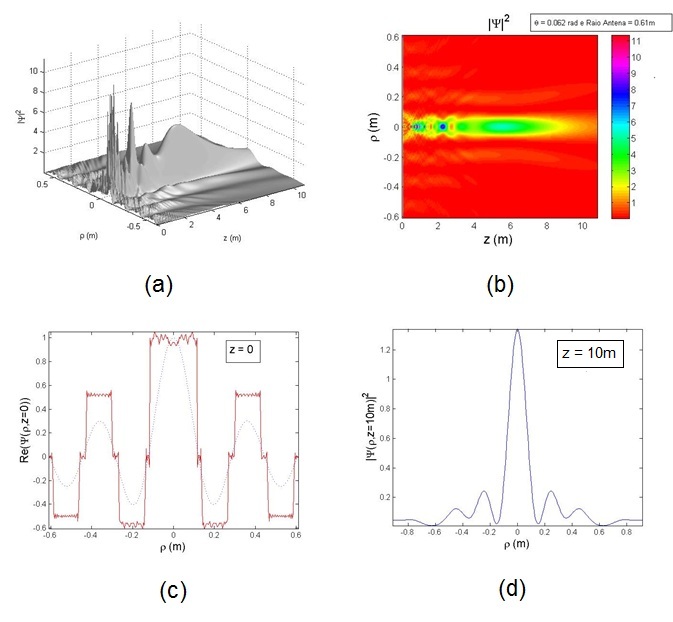}}
\end{center}
\caption{Campo emanato dal Prototipo 2. \ Spiegazioni nel testo.} \label{fig7}
\end{figure}

\h Qui l'idea \`e stata di aumentare l'intensit\`a del campo negli slits (eccetto quello
dell'apertura centrale), dato che, cos\`{\i} facendo, il raggio dello spot si
mantiene invariato, ma la distribuzione di intensit\`a del fascio in $\rho=0$ (ossia, l'intensit\`a
centrale dello spot) diventa pi\`u omogenea che non nel caso del Prototipo 1. \ Inoltre,
\`e facile vedere che migliora sostanzialemente l'intensit\`a dello spot in $z=10$ m.

\

\

\

\h \textbf{Prototipo numero 3}: Di nuovo, manteniamo inalterate le dimensioni dell'antenna, e cambiamo
solamente il valore dei campi uniformi che ``illuminano" ciascuno slit. \ Applichiamo ora il
medesimo modulo di ampiezza (ovvero, la stessa intensit\`a!) a tutti gli slits!, mutando solo
il suo segno, che sar\`a alternativamente positivo o negativo passando da uno slit
all'altro. In altre parole, cambiamo solo la fase [di un banale valore $\pi$] passando da uno
slit al vicino.

\h Numericamente avremo: $\Psi_0 = 1$ a.u., \ $\Psi_1 = -1$ a.u., \ $\Psi_2 = 1$ a.u., \
$\Psi_3 = -1$ a.u.

\h Le Figure 8 mostrano il campo emanato dal Prototipo 3.

\newpage

\begin{figure}[!h]
\begin{center}
 \scalebox{1.1}{\includegraphics{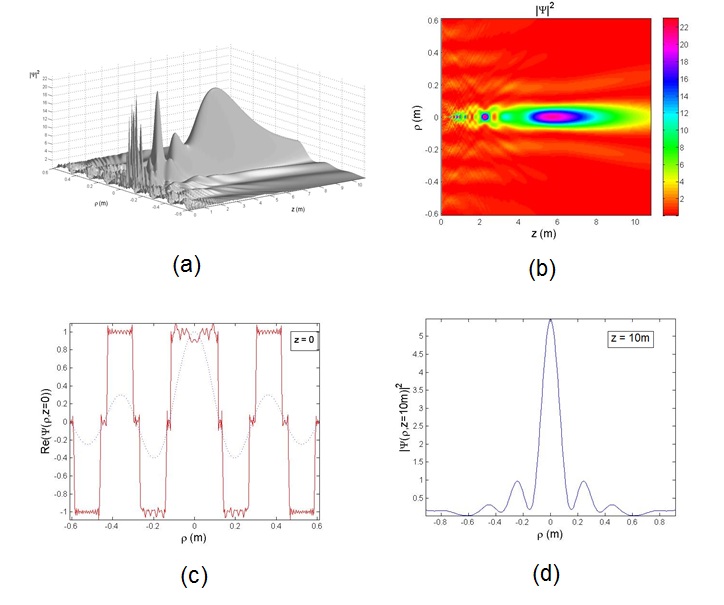}}
\end{center}
\caption{Campo emanato dal prototipo 3. \ Spiegazioni nel testo.} \label{fig8}
\end{figure}

\h Ancora una volta aumentiamo l'intensit\`a del campo negli slits laterali, dato che cos\`{\i}
si osserva che il raggio dello spot apparentemente non subisce alterazioni, mentre
si ha un aumento di omogeneit\`a dell'intensit\`a sull'asse, insieme con un {\em aumento}
dell'intensit\`a dello spot in $z=10$ m rispetto al Prototipo 1.

\h E' pure interessante notare che, nonostante la distribuzione omogenea dell'intensit\`a
negli slits, il campo emanato diventa rapidamente un fascio con campo concentrato
intorno a $\rho=0$. \ Ci\`o \`e dovuto al cambiamento alternato di fase (ogni volta della
quantit\`a $\pi$)
passando da uno slit al successivo.

\newpage

{\bf 4. -- Achnowledgmnents}\\

The first two authors acknowledge partial support from CATES, FAPESP and INFN. One of them
(ER) thanks DMO/FEEC/UNICAMP for kind hospitality.

\

\

\

{\bf 5. -- An essential Bibliography}\\

--- H.E.H.Figueroa, M.Z.Rached and E.Recami (editors):
{\em Localized Waves}
(J.Wiley; New York, Jan.2008); book of 386 pages.\hfill\break

--- E.Recami and M.Z.Rached: ``Localized Waves: A not-so-short
Review",\hfill\break {\em Advances in Imaging \& Electron Physics
(AIEP)} 156 (2009) 235-355\hfill\break [121 printed
pages], {\bf and references therein}.\hfill\break

--- M.Z.Rached, E.Recami and M.Balma: ``A simple and efficient analytical approach
to the description of optical beams truncated by finite apertures", to be submitted
for pub.

\end{document}